\documentstyle[preprint,eqsecnum,aps,epsfig]{revtex}
\tightenlines
\begin{document}
\draft
\preprint{\vbox{\hbox{UL--NTZ 10/97}\hbox{HUB--EP--97/24}}} 
\title{
Where the electroweak phase transition ends}
\author{
  M.~G\"urtler$^1$\thanks{guertler@tph204.physik.uni-leipzig.de}}
  \address{Institut f\"ur Theoretische Physik, Universit\"at Leipzig, Germany}
\author{
  E.--M.~Ilgenfritz$^2$\thanks{ilgenfri@pha1.physik.hu-berlin.de}}
  \address{Institut f\"ur Physik, Humboldt-Universit\"at zu Berlin, Germany}
\author{
  A.~Schiller$^1$\thanks{schiller@tph204.physik.uni-leipzig.de}} 
  \address{Institut f\"ur Theoretische Physik, Universit\"at Leipzig, Germany}
\date{June 11, 1997}
\maketitle
\begin{abstract}
  We give a more precise characterisation of the end of the electroweak phase
  transition in the framework of the effective $3d$ $SU(2)$--Higgs lattice
  model than has been given before.  The model has now been simulated at gauge
  couplings $\beta_G=12$ and $16$ for Higgs masses $M_H^*=70, 74, 76$ and
  $80$~GeV up to lattices $96^3$ and the data have been used for reweighting.
  The breakdown of finite volume scaling of the Lee--Yang zeroes indicates the
  change from a first order transition to a crossover at
  $\lambda_3/g_3^2=0.102(2)$ in rough agreement with results of
  Ref.~\cite{karsch96} at $\beta_G=9$ and smaller lattices.  The infinite
  volume extrapolation of the discontinuity $\Delta \langle \phi^+ \phi
  \rangle /g_3^2$ turns out to be zero at $\lambda_3/g_3^2=0.107(2)$ being an
  upper limit. We comment on the limitations of the second method.
\end{abstract}

\narrowtext

\section{Introduction}
\label{sec:intro}

During the last couple of years, big effort has been invested to study the
properties of the first order phase transition that the standard model was
expected to undergo at high temperature (for reviews see \cite{lat94-96}).
The motivation was to explore the phenomenological viability of the generation
of the baryon asymmetry of the universe at this transition.

The perturbative evaluation of the phase transition is prevented by infrared
divergences in the so--called symmetric phase.  Lattice Monte Carlo studies of
the $4$ dimensional $SU(2)$--Higgs theory \cite{4d,desy} have been done so far
for relatively large lattice spacings (neglecting the U(1) gauge group and the
fermionic content of the theory).  Another approach is based on the concept of
dimensional reduction \cite{dimred}.  One maps the theory (with or without
fermions) onto a $3$ dimensional $SU(2)$--Higgs model containing all infrared
problems of the full theory and can investigate this effective theory by Monte
Carlo simulations \cite{KajantieNP96,3d,wirNP97,karsch96} with much less
effort. Later, the $U(1)$ gauge group has been included into this approach,
too \cite{hki_u1}.

BAU generation at the phase transition of the standard model is ruled out
already, and the primary interest has shifted to extensions of the standard
model.  Nevertheless, the standard variant remains interesting in order 
\begin{enumerate}
\item [({\it i})]to understand methods like dimensional reduction, validity of
  perturbation theory etc. in the realm of extremely weakly first order phase
  transitions and
\item [({\it ii})] to understand in general terms the physics in the strongly
  coupled high temperature phase of gauge--matter systems.
\end{enumerate}

The present paper belongs to the first group of studies. We try to shed light
on the question for which Higgs mass the first order transition ceases to
exist and what replaces it at slightly higher Higgs mass.  Analytical work has
already addressed this problem. In \cite{reuter93} it has been claimed that
the transition between the broken and the symmetric phase can only be of first
order or a smooth crossover. Within the same average action approach it has
been made more precise later \cite{bergerhoff} that the first order transition
ends at a Higgs mass of about $80$ GeV and the transition is replaced by a
unique strongly interacting phase.  A similar conclusion has been drawn from a
renormalisation group study of the electroweak phase transition
\cite{tetradis}.  Analysing gap equations a similar critical Higgs mass has
been pointed out in Ref. \cite{buchmueller95}.

Recently, $3d$ Monte Carlo studies \cite{kajantieprl96,karsch96} have
investigated the volume dependence of the susceptibility of the Higgs
condensate. These studies gave support for the claim that the transition turns
into a smooth crossover for large Higgs masses. An attempt to determine the
value of the upper critical Higgs mass has been performed in \cite{karsch96}.
It was based on an analysis of the volume dependence of the Lee--Yang zeroes
\cite{yang}, but for a relatively large lattice spacing.  The exploration of
critical behaviour in lattice gauge theories using Lee--Yang zeroes has become
a frequently used tool nowadays. A good guide to the basic applications can be
found in \cite{marinari}.

The only $4d$ study to determine the critical Higgs mass region has been
presented in \cite{aoki}, however with a temporal extent of only $N_t=2$ and
an exploratory scan of the Higgs self coupling (corresponding to different
Higgs masses).

In our present study we use Monte Carlo simulations of the three dimensional
theory in order to find the critical Higgs mass, employing two different types
of analysis.  The first is plainly to look for the Higgs mass where the jump
of the scalar condensate (which is proportional to the latent heat) vanishes.
The second method is based on an analysis of the Lee--Yang zeroes of the
partition function, whose finite volume behaviour changes with the character
of the transition and is able to characterise the change of first order into
crossover.

\section{The model and numerical techniques}
\label{sec:tech}

The lattice $3d$ $SU(2)$--Higgs model is defined by the action
\begin{eqnarray}
  S &=& \beta_G \sum_p \big(1 - {1 \over 2}{\mathrm tr} U_p \big) - \beta_H
  \sum_{x,\alpha}E_{x,\alpha} \nonumber \\
 & &+ \sum_x \big( \rho_x^2 + \beta_R
  (\rho_x^2-1)^2 \big)
   \label{eq:latt_action}
\\
E_{x,\alpha}&=&{1\over 2}{\mathrm tr} (\Phi_x^+ U_{x, \alpha} \Phi_{x +
    \alpha})
   \label{eq:latt_action2}
\end{eqnarray}
(summed over plaquettes $p$, sites $x$ and directions $\alpha$), with the
gauge coupling $\beta_G$, the lattice Higgs self--coupling $\beta_R$ and the
hopping parameter $\beta_H$. The gauge fields are represented by unitary $2
\times 2$ link matrices $U_{x,\alpha}$ and the Higgs fields are written as
$\Phi_x = \rho_x V_x$. $\rho_x^2= \frac12{\mathrm tr}(\Phi_x^+\Phi_x)$ is the
Higgs modulus squared, $V_x$ an element of the group $SU(2)$, $U_p$ denotes
the $SU(2)$ plaquette matrix. For shortness, we characterise as in
\cite{wirNP97} the Higgs self--coupling by an approximate Higgs mass $M_H^*$
defined through
\begin{equation}
  \label{eq:mh*}
  \beta_R=\frac {\lambda_3}{g_3^2}\frac{\beta_H^2}{\beta_G}
  =
  \frac 18{\left(\frac{M_H^*}{80\ {\mathrm GeV}}\right)}^2
  \frac{\beta_H^2}{\beta_G},
\end{equation}
where $\lambda_3$ and $g_3$ are the dimensionful quartic and gauge couplings
of the corresponding $3d$ continuum model, respectively.  Both couplings are
renormalisation group invariants. The $3d$ continuum model is furthermore
characterised by the renormalised mass $m_3(g_3^2)$ taken at the scale
$\mu_3=g_3^2$. To study the continuum limit of the lattice model at given
$M_H^*$ (along the line of constants physics of the $3d$ continuum theory) one
has to keep the coupling ratios $\lambda_3/g_3^2$ and $m_3(\mu_3=g_3^2)/g_3^2$
fixed.

Letting increase the gauge coupling $\beta_G$ at fixed $\lambda_3/g_3^2$ along
the critical line dividing the high temperature and Higgs phase
($m_3(g_3^2)/g_3^2$ fixed and near to zero) permits to perform the continuum
limit according to the relation
\begin{equation}
  \label{eq:betag}
  \beta_G=\frac 4 {a g_3^2}
\end{equation}

We study the phase transition driven by the hopping parameter $\beta_H$. The
Monte Carlo simulations are performed at $\beta_G=12$ and $\beta_G=16$ for
different $M_H^*$, ranging from $70$ to $80$~GeV at several values of
$\beta_H$ on cubic lattices of the size $L^3$ (see Table~\ref{tab:statistics}
for parameters and statistics).  The simulations have been performed at the
DFG Quadrics computer QH2 in Bielefeld and on the CRAY-T90 of the HLRZ
J\"ulich, some data have been collected on a Q4 Quadrics in J\"ulich. For the
update we used the same algorithm as described in \cite{wirNP97} which
combines Gaussian heat bath updates for the gauge and Higgs fields with
several reflections for the fields to reduce the autocorrelations.  All
thermodynamical bulk quantities are measured after each such combined sweep.
One combined sweep with bulk measurements takes $3.5$ sec on a $96^3$ lattice
on the QH2 parallel computer.

In the search for the phase transition the space averaged square of the Higgs
modulus
\begin{equation}
  \label{eq:rho2}
  \rho^2=\frac{1}{L^3}\sum_x \rho_x^2
\end{equation}
is used, $\langle\rho^2 \rangle$ denotes averaging over the Monte Carlo
measurements.

In our analysis we have used the Ferren\-berg--Swendsen method \cite{ferswen}.
Note that at fixed $M_H^*$ (and $\beta_G$) the Higgs self--coupling $\beta_R$
is quadratic in $\beta_H$ (see Eq. \ref{eq:mh*}). The reweighting has to be
performed by histogramming in two parts of the action. The partition function
is represented as
\begin{eqnarray}
  \label{eq:sos_from_dos}
  \lefteqn{ Z(L,\beta_H)=}  \nonumber \\
  && \int dS_1 dS_2 D_L(S_1,S_2) 
  \exp\left(L^3 (\beta_H S_1 -\beta_H^2 S_2)\right)\ ,
\end{eqnarray}
\begin{equation}
  S_1=3 E_{\mathrm link}\ ,\ \ \  S_2= \frac{\lambda_3} 
  {g_3^2 \beta_G}(\rho^4-2 \rho^2)
\end{equation}
where the density of states $D_L(S_1,S_2)$ is approximated by the histogram
produced by multihistogram reweighting of all available data for given
$\beta_G$ and $L^3$ (see Table~\ref{tab:statistics}).  Having a good estimator
of the density of states $D_L(S_1,S_2)$ from a sufficient number of simulation
points we are able to
\begin{itemize}
\item interpolate in $\beta_H$ at fixed $M_H^*$ to localise the phase 
transition,
\item interpolate in $M_H^*$ in order to find the critical Higgs mass,
\item extrapolate to complex $\beta_H$ to study Lee--Yang zeroes.
\end{itemize}

Finally, all considered quantities are translated into physical units.  This
allows to combine results obtained for different $\beta_G$ and gives a check
to what extent the continuum limit is reached. The size of the lattice in
continuum length units ({\it i.e.} the inverse $3d$ gauge coupling $g_3^2$) is
given by the expression
\begin{equation}
  \label{eq:length}
  l g_3^2=L a g_3^2= 4L/\beta_G   
\end{equation}
and the jump of the quadratic scalar condensate is in the corresponding mass 
units
\begin{equation}
  \label{eq:condensate}
\frac{\Delta \langle\phi^+ \phi\rangle}{g_3^2} = 
\frac18 \beta_G \beta_H \Delta \langle \rho^2 \rangle \ .
\end{equation}
Here $\Delta \langle \rho^2 \rangle =\langle \rho^2_{\mathrm b}
\rangle-\langle \rho^2_{\mathrm s}\rangle$ denotes the difference of the
lattice quadratic scalar condensates measured at the pseudo--critical hopping
parameter between the broken ($\langle \rho^2_{\mathrm b}\rangle$) and
symmetric ($\langle \rho^2_{\mathrm s}\rangle$) phases, respectively.

\section{The behaviour of the latent heat with increasing Higgs mass}
\label{sec:latheat}

A non--vanishing latent heat $\Delta \epsilon$ is one of the characteristics
for a first order phase transition. In our model the latent heat is
proportional to the jump of the quadratic scalar condensate $\Delta
\langle\phi^+\phi\rangle$ \cite{KajantieNP96}.  The proper identification of
the scalar condensate discontinuity becomes increasingly demanding near to the
end of the first order transition. The correlation length grows beyond the
size of the system under study, in particular if the endpoint is a critical
point. There can be an apparent metastability on a finite torus which delays
the approach to the thermodynamical limit.

In this section we identify the end of the phase transition with the point in
the $\beta_H$--$M_H^*$ plane where the discontinuity vanishes.

\begin{figure}
  \centering
  \epsfig{file=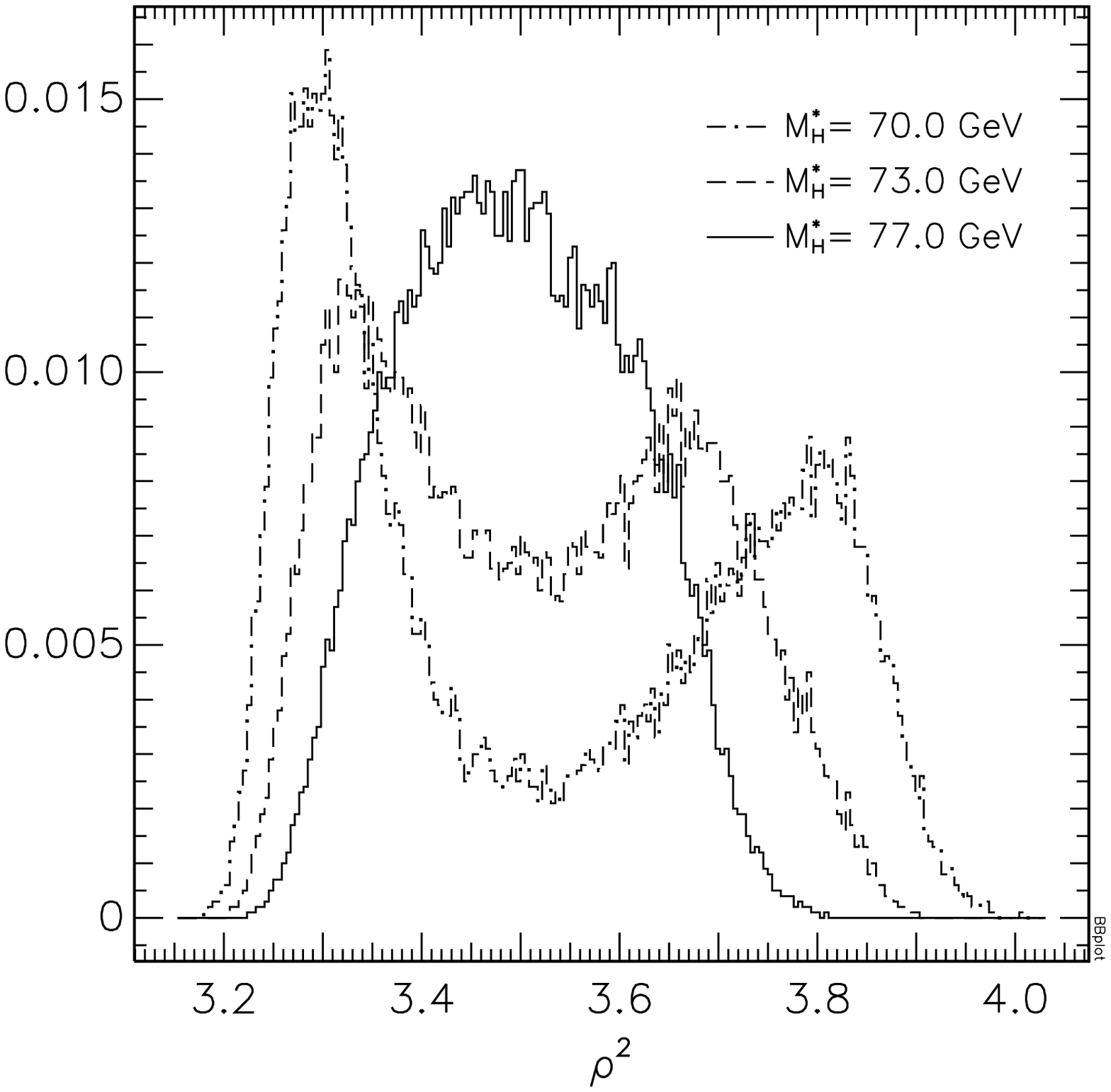,width=86mm}
  \caption{Histograms of $\rho^2$ for different $M_H^*$ at the respective 
    pseudo--critical $\beta_H$ (defined by the minimum of the Binder cumulant)
    for a $80^3$ lattice, $\beta_G=12$}
  \label{fig:histograms}
\end{figure}

The minimum of the Binder cumulant 
\begin{equation}
  \label{eq:binder_cum} 
  B_{\rho^2}(L,\beta_H)= 1 - {\frac{{\langle
        ({\rho^2})^4 \rangle} }{{\ 3 {\langle ({\rho^2})^2 \rangle}^2}}}
\end{equation}
and the maximum of the susceptibility of $\rho^2$
\begin{equation}
  \label{eq:suscept} 
  C_{\rho^2}(L,\beta_H) = \langle ({\rho^2}) ^2
  \rangle  - \langle {\rho^2} \rangle ^2
\end{equation}
are chosen to define pseudo--critical values of the hopping parameter
$\beta_H$.  The jumps in $\Delta \langle \rho^2 \rangle$ are extracted from
the peaks of the histograms reweighted to these values of $\beta_H$.  The
$\rho^2$ histograms at the respective pseudo--critical couplings show how the
discontinuity decreases with increasing $M_H^*$. The gap between the peaks is
more and more filled, and the distance between them becomes smaller
(Fig.~\ref{fig:histograms}).

At any Higgs mass $M_H^*$ we attempt to perform the thermodynamical limit of
$\Delta \langle \rho^2 \rangle$ by assuming the finite size corrections to
follow an inverse cross sectional law (suggested by the behaviour of the Potts
model in $2$ dimensions \cite{borgs})
\begin{equation}
  \label{eq:finite}
  \left|\Delta \langle \phi^+ \phi \rangle_\infty-\Delta \langle \phi^+ \phi
    \rangle_l\right| \propto 1/l^2\,.
\end{equation}

\begin{figure}
  \centering 
  \epsfig{ file=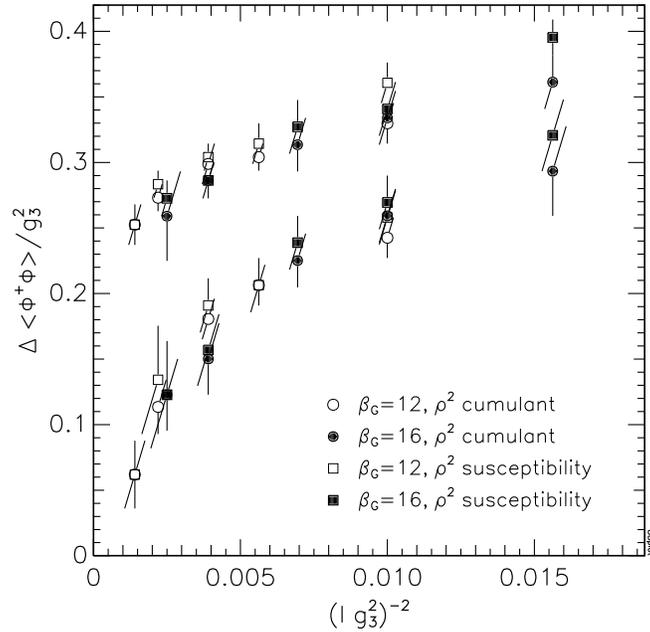, angle=0,width=86mm}
  \caption{
    Quadratic Higgs condensate jump $\Delta \langle \phi^+ \phi \rangle/g_3^2$
    as function of inverse physical length squared, upper data correspond to
    $M_H^*=70$~GeV, lower to $M_H^*=76$~GeV}
  \label{fig:all_vol70+76}
\end{figure}

Fig.~\ref{fig:all_vol70+76}
shows the different infinite volume extrapolation of $\Delta\langle \phi^+
\phi \rangle/g_3^2$ at $M_H^*=70$ and $76$~GeV. We have used two criteria
(minimum of the Binder cumulant and maximum of the susceptibility for the
lattice quadratic Higgs condensate) to determine the pseudo--critical
$\beta_H$. One observes that the data for different $\beta_G$ cluster along
one curve within their errors and do not show a significant dependence on the
lattice spacing $a$.  Therefore, we conclude that the quadratic Higgs
condensate jump at the measured $\beta_G$ values is already sufficiently near
to the continuum limit.  Obviously the latent heat at $M_H^*=76$~GeV has a
vanishing thermodynamical limit.  At the larger Higgs mass the assumed $1/l^2$
scaling compatible with the vanishing limit of the condensate sets in only for
the largest considered volumes.

\begin{figure}
  \centering 
  \epsfig{file=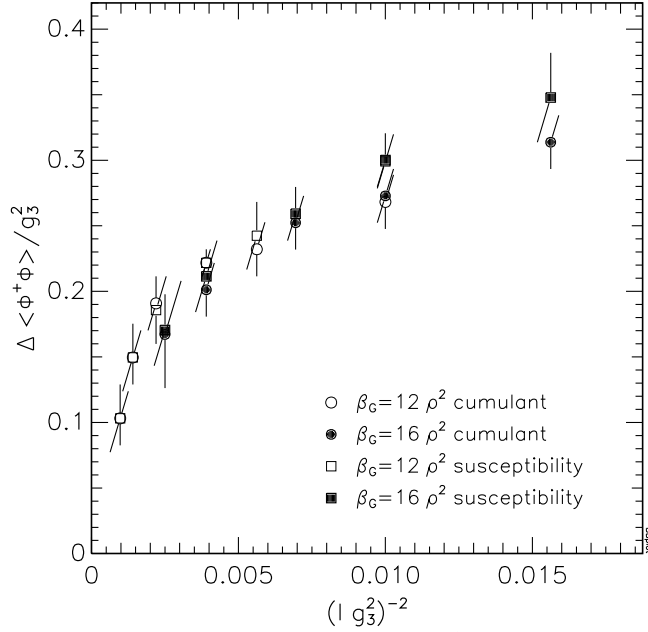, angle=0,width=86mm}
  \caption{
    Quadratic Higgs condensate jump $\Delta \langle \phi^+ \phi \rangle/g_3^2$
    as function of inverse physical length squared at $M_H^*=74$~GeV}
  \label{fig:all_vol74}
\end{figure}

This feature becomes even more pronounced at intermediate Higgs mass,
$M_H^*=74$~GeV.  Fig. \ref{fig:all_vol74} shows the volume dependence of the
condensate jump for $M_H^*=74$~GeV. This picture includes, besides reweighted
data, original simulations at that Higgs mass for lattices up to $96^3$ at
$\beta_G=12$. The onset of the $1/l^2$--scaling is delayed to lattices not
smaller than $80^3$ (for $\beta_G=12$). If only those data are considered the
latent heat is consistent with zero.

\begin{figure}
  \centering
  \epsfig{file=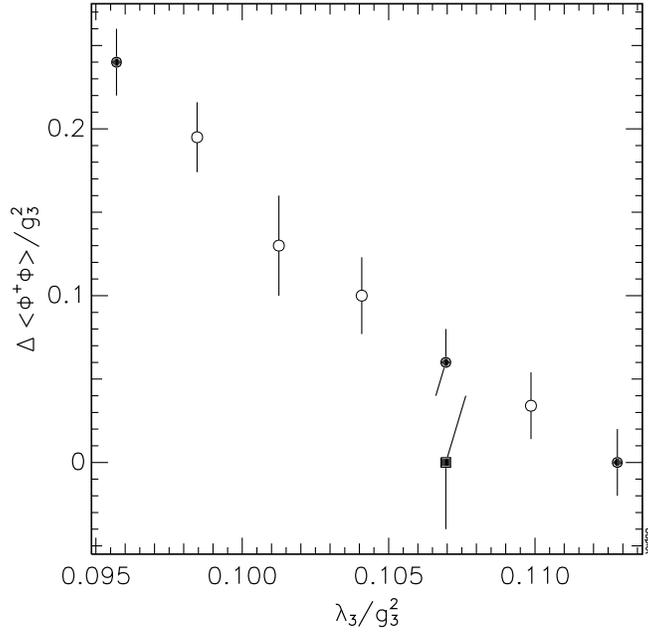,width=86mm}
  \caption{
    Infinite volume discontinuity $\Delta \langle \phi^+\phi \rangle/g_3^2$
    shown vs.  $\lambda_3/g_3^2$. Filled symbols mark the Higgs masses
    $M_H^*=70$, $74$ and $76$~GeV where data have been taken, open symbols
    denote results from FS interpolation. The isolated lower data point at
    $M_H^*=74$ GeV refers to an infinite volume extrapolation including only
    $80^3$ and $96^3$ lattices as described in the text.}
  \label{fig:thd_lim}
\end{figure}

The summary of the extrapolations to the thermodynamical limit is collected in
Fig.~\ref{fig:thd_lim}.  For the extrapolation according to Eq.
(\ref{eq:finite}) we have used the results for the discontinuity at $(l
g_3^2)^{-2} < 0.003$ which correspond to lattices $\ge 64^3$ for $\beta_G=12$
and $80^3$ for $\beta_G=16$.  For $\lambda_3/g_3^2\approx 0.107$
($M_H^*=74$~GeV) a different extrapolation is shown, lying below the general
trend, which is compatible with zero at that Higgs mass. This extrapolation
takes into account only the two largest volumes at $\beta_G=12$. The
uncertainty reflects the scattering of slopes of the straight line
interpolation.

Concerning the two--parameter multihistogram extrapolation we can report that
the purely interpolated histograms at $M_H=74$~GeV near to the endpoint are in
reasonable agreement with histograms at that mass supported by actual
simulations. However, at that mass we are too near to the endpoint, such that
simulations at still larger lattices are necessary in order to estimate the
correct thermodynamical limit.
 
We conclude that the critical coupling, if defined by vanishing latent heat
(vanishing jump of the quadratic scalar condensate), is bounded from above as
\begin{equation}
  \label{eq:l3g32_first}
  \lambda_{3 {\mathrm crit}}/g_3^2 < 0.107\  .
\end{equation}
This bound is somewhat larger than the critical coupling given in
\cite{karsch96}. It could be tempting to explain this difference to the
somewhat smaller $\beta_G=9$ in their paper.

In this section we have assumed an early continuum limit by plotting data from
different $\beta_G$ as a function of the physical lattice volume. The data
were compatible with each other within the errors. The account for finite $a$
corrections of expectation values in the thermodynamical limit could be
performed (if necessary) along the line of Ref. \cite{moore}.

\section{Lee--Yang zeroes near the critical Higgs mass}

In this section we will determine the critical Higgs mass by analysing the
position of the Lee--Yang zeroes in the complex $\beta_H$ plane and their
motion with increasing size of the finite lattice system.

Phase transitions correspond to non--analytical behaviour of the infinite
volume free energy density as function of couplings which normally are
real--valued.  This is signalled by zeroes in the complex plane (to which the
relevant coupling constant is extended) of the partition function of finite
systems.  If there exists a phase transition driven by this coupling some of
these zeroes cluster, in the thermodynamical limit, along lines that pinch the
real axis. This prevents the analytic continuation along the real axis
corresponding to that coupling.
 
We sketch here the motion of the most important zeroes with increasing volume.
Neglecting interface tension effects the partition function at the transition
point is given by the contributions of the two phases
\begin{equation}
Z = Z_s + Z_b= e^{-L^3 f_s} + e^{-L^3 f_b}
\end{equation}
where $f_{s(b)}$ denotes the lattice free energy per site of the so--called
symmetric (broken) phase.  The free energy density can be expanded around the
real--valued pseudo--critical coupling $\beta_{Hc}$
\begin{eqnarray}
  f_{s,b}(\beta_H)&=&f(\beta_{Hc}) + \langle E_{s,b}(\beta_{Hc}) \rangle
  (\beta_{H}-\beta_{Hc}) \nonumber \\ 
                  &&+{\cal O}((\beta_H-\beta_{Hc})^2)\ .
\end{eqnarray} 
To obtain $\langle E \rangle$ we use the action in the form 
\begin{equation}
  S= S_0 - 3 \beta_H L^3 E_{\mathrm link}+\beta_R^2 L^3 (\rho^4 -2 \rho^2)
\end{equation}
with $\rho^2$ defined in Eq.~(\ref{eq:rho2}) and 
\begin{equation}
  \label{eq:qqq}
  E_{\mathrm link}=\frac{1}{3L^3}\sum_{x,\alpha}E_{x,\alpha}\ ,\ \ \
  \rho^4=\frac{1}{L^3}\sum_x \rho_x^4 \ .
\end{equation}
Taking into account that the Higgs self--coupling is (for given $M_H^*$)
quadratic in $\beta_H$ we find
\begin{equation}
  \label{eq:e_betahc}
  \langle E(\beta_{Hc})\rangle=-3\langle E_{\mathrm link}\rangle+2
  \frac{\beta_{Rc}}{\beta_{Hc}}\left( \langle \rho^4 \rangle - 2 \langle
    \rho^2 \rangle \right)
\end{equation}
where $\langle \dots \rangle$ denotes the Monte Carlo average as before.

Using the decomposition at $\beta_{Hc}$
\begin{equation}
  \langle E_{s(b)} \rangle =  \langle E \rangle  \mp \frac 12 \Delta 
  \langle E \rangle  \ ,
\end{equation}
the partition sum behaves as
\begin{equation}
  Z 
  \propto
  \cosh \left[ \frac{ \Delta \langle E \rangle }{2} L^3
 (\beta_{H}-\beta_{Hc})\right] \ .
\end{equation}
For the complex coupling $ \beta_H= {\mathrm Re} \beta_H + i ~{\mathrm Im}
\beta_H $ we obtain in this approximation that the zeroes of the complex
partition function $Z$ are located at ($n$ are integers)
\begin{eqnarray}
  {\mathrm Im} \beta_H^{(n)} &=&
  \frac{2 \pi }{ L^3 \left|  
      \Delta \langle E \rangle \right|} \left(n-\frac{1}{2}\right) \ , \\
  {\mathrm Re} \beta_H&=& \beta_{Hc} \ .
\end{eqnarray}
For a volume independent $\Delta \langle E \rangle$ the imaginary part of the
hopping parameter at the position of the zeroes would scale with the inverse
volume. For infinite volume the zeroes become dense and prevent analytic
continuation of $Z$ beyond $\beta_{Hc}$.

Taking into account Eq.~(\ref{eq:e_betahc}) and using the identity for the
condensate jumps \cite{wirNP97}
\begin{eqnarray}
  - 3 \beta_{Hc} \ \Delta \langle {E_{link}} \rangle + (1&-&2\beta_{Rc})
  \Delta \langle {\rho^2} \rangle \nonumber \\  
 + 2 \beta_{Rc} \ \Delta \langle
  {\rho^4} \rangle&=& 0 
\end{eqnarray}
one easily finds
\begin{equation}
  - \Delta \langle E \rangle 
  = 
  \frac{1+2 \beta_{Rc}}{\beta_{Hc}} \Delta \langle \rho^2 \rangle
   \ .
\end{equation}
Therefore, for the phase transition still being first order one expects the
approximate relation between the imaginary part of the leading zeroes in the
complex hopping parameter plane (with index $n$) and the Higgs condensate
discontinuity
\begin{equation}
  {\mathrm Im} \beta_H^{(n)}= 
  \frac{2 \pi \beta_{Hc}}{L^3 (1+2 \beta_{Rc})
    \Delta \langle \rho^2 \rangle} \left(n-\frac{1}{2}\right) \ .
  \label{eq:zero_as_fct_of_rho2}
\end{equation}

Since $ \Delta \langle \rho^2 \rangle$ itself depends on the size of the
lattice (as discussed in section~\ref{sec:latheat}) the simple $1/L^3$
behaviour for $ {\mathrm Im} \ \beta_H$ is modified and can be expected only
asymptotically.

An analysis of the first Lee--Yang zero in the crossover region of the $3d$
$SU(2)$--Higgs model has been carried out recently in Ref.~\cite{karsch96}.
Here we are interested to discuss in more detail the change from first order
transition to a crossover behaviour at the critical Higgs mass.  

As usual, the partition function has to be analytically continued into the
complex plane as function of the complex hopping parameter $\beta_H$ near to
the real pseudo--critical coupling $\beta_{Hc}$.  This can be done by
reweighting (\ref{eq:sos_from_dos}).  Since $\beta_H$ is complex, the action
$S$ and consequently $Z$ become complex, too.  The zeroes of $Z$ are found
numerically using the Newton--Raphson method for solving simultaneously
${\mathrm Re} Z=0$ and ${\mathrm Im} Z=0$. To estimate the accuracy of the
position of the zeroes in the complex plane we have calculated them using only
the half data sample.

\begin{figure}
  \vspace{1cm}
  \centering
  \epsfig{file=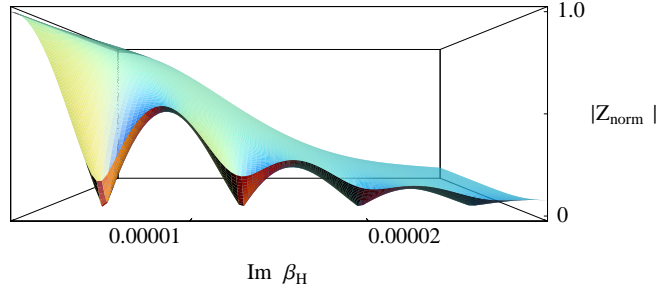,width=86mm,angle=0}
  \caption{
    $3d$ view of $\left|Z_{\mathrm{norm}}\right|$ near to the first zeroes at
    $\beta_G=12$, $80^3$ and $M_H^*=70$~GeV}
  \label{z2_12_80mod}
\end{figure}

In Fig.~\ref{z2_12_80mod} the modulus of the complex partition function
$\left|Z_{\mathrm norm}\right|$ in the neighbourhood of the pseudo--critical
hopping parameter $\beta_{Hc}$ is shown.  $Z_{\mathrm norm}$ has been used for
clarity. This means that $Z(\beta_H)$ is divided, for each complex $\beta_H$,
by its (real) value at ${\mathrm Re} \beta_H$, $Z({\mathrm Re} ~\beta_H)$.
The figure represents a lattice of size $80^3$ at $\beta_G=12$ for a Higgs
mass $M_H^*=70$~GeV where the transition is still clearly first order
\cite{wirNP97}. The normalised $\left| Z_{\mathrm norm}\right|$ approaches zero in
the clearly distinct minima.
 
\begin{figure}
  \centering  
  \epsfig{file=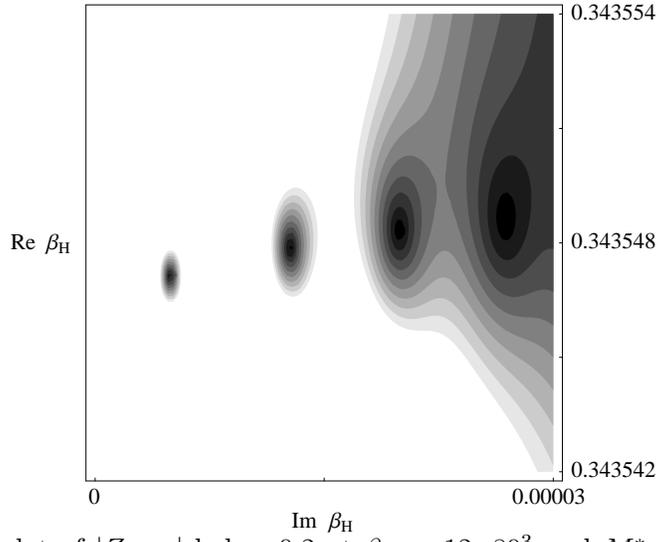, width=86mm,angle=0}
  \caption{
    Contour plot of $\left|Z_{\mathrm{norm}}\right|$ below 0.2 at
    $\beta_G=12$, $80^3$ and $M_H^*=70$~GeV with height differences of 0.02 }
  \label{fig:cont70}
\end{figure}

\begin{figure}
  \centering 
  \epsfig{file=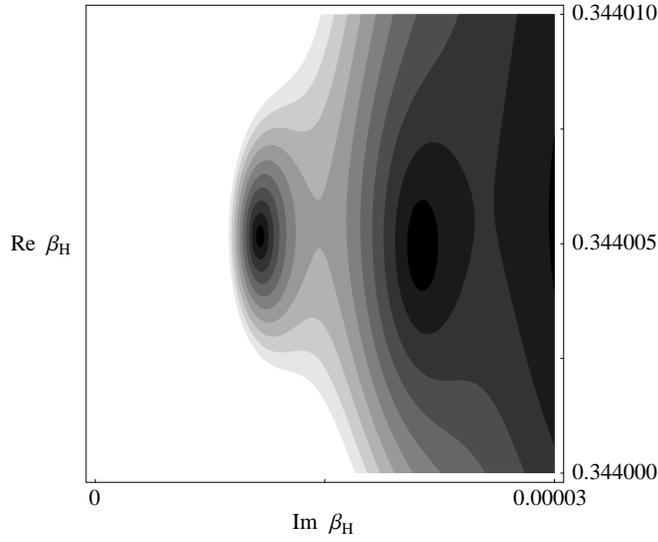, width=86mm,angle=0}
  \caption{
    Same as Fig.~6 at $\beta_G=12$, $80^3$ and $M_H^*=76$~GeV }
  \label{fig:cont76}
\end{figure}

The difference in the pattern of the leading complex zeroes is demonstrated in
Figs.~\ref{fig:cont70} and \ref{fig:cont76} referring to $M_H^*=70$ and $76$
GeV for the same lattice size $80^3$ and lattice gauge coupling $\beta_G=12$.
Each figure shows a part of the strip $0 \le \mathrm{Im} \beta_H \le 3 \times
10^{-4}$ along the real axis where the leading Lee--Yang zeroes are located at
the respective Higgs mass.  For the larger Higgs mass the normalised modulus
decreases much faster with increasing ${\mathrm Im}\beta_H$, less zeroes are
localised inside the strip and the funnels which form the
$\left|Z_{\mathrm{norm}}\right|$ landscape at the locations of the Lee--Yang
zeroes become less steep. The zeroes move away from the real axis with
increasing Higgs mass.  Notice that only for the lower Higgs mass the pattern
approximately follows the $n$ dependence given in
Eq.~(\ref{eq:zero_as_fct_of_rho2}).  This is a hint for an inherently
different behaviour of the model at these selected Higgs masses. To answer the
question whether this difference shows the vanishing of the phase transition
we investigate the zeroes in the thermodynamical limit.

\begin{figure}
  \centering 
  \epsfig{file=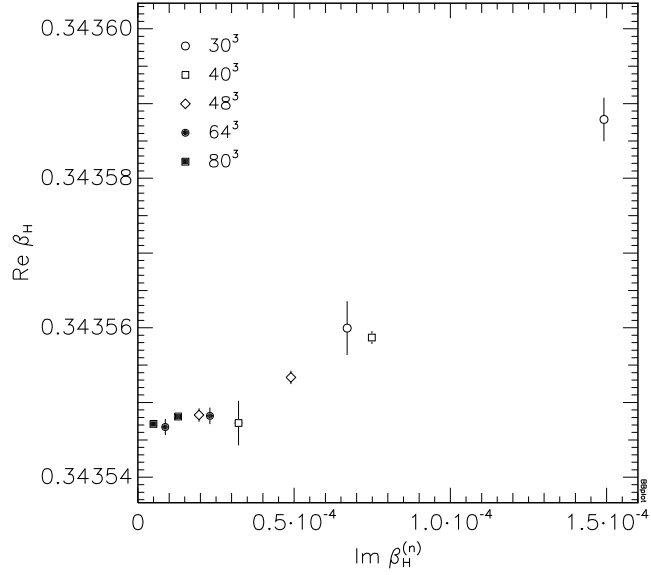, width=86mm,angle=0}
  \caption{First two zeroes at $\beta_G=12$ and $M_H^*=70$~GeV for
    different lattice sizes}
  \label{zero_12_70}
\end{figure}

In Fig.~\ref{zero_12_70} the first two zeroes of different lattice volumes are
collected. There is a tendency of the zeroes to move to larger ${\mathrm Re}
\beta_H$ with decreasing lattice volume and increasing index of the zero (for
low $n$). The first tendency corresponds with the fact that the maximum of the
link susceptibility gives a pseudo--critical $\beta_H$ which approaches the
infinite volume limit $\beta_{Hc}$ from above \cite{wirNP97}.

To extract information about the endpoint of the transition we fit the
imaginary part of the first zero for each available physical length $l$
(Eq.~\ref{eq:length}) according to
\begin{equation}
  \label{eq:zero_fit}
  {\mathrm Im} \beta_H^{(1)}=C (lg_3^2)^{-\nu}+R \ .
\end{equation}
A positive $R$ in Eq.~(\ref{eq:zero_fit}) should indicate that the first zero
does not approach anymore the real axis in the thermodynamical limit (as
required for a phase transition) and our first order transition has turned
into a crossover.

\begin{figure}
  \centering
  \epsfig{file=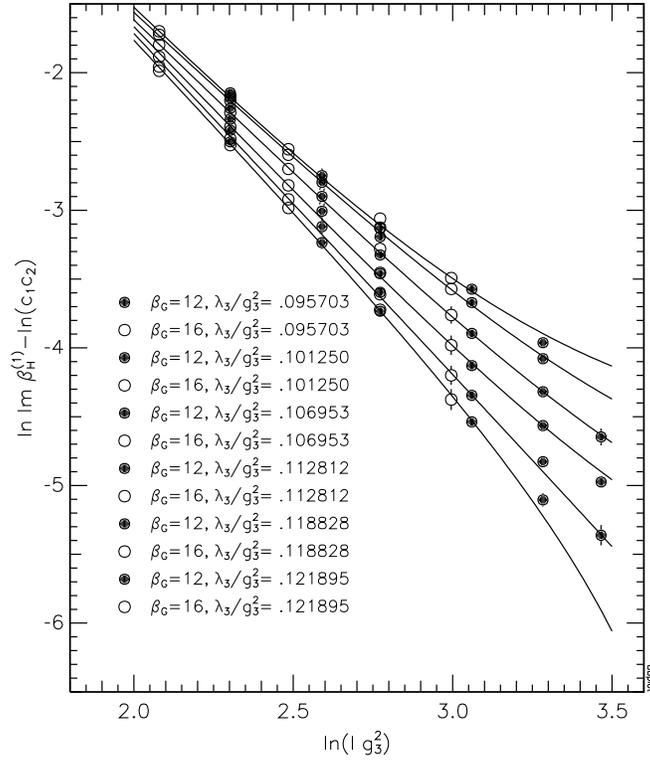, width=86mm,angle=0}
  \caption{
    Logarithm of the imaginary part of first zeroes at different
    $\lambda_3/g_3^2$ vs. logarithm of the physical length $\ln (l g_3^2)$
    together with the fit described in the text}
  \label{fig:log_im_vs_log_L}
\end{figure}

We assume that for equal physical volume the imaginary part of the Lee--Yang
zeroes shows a universal behaviour.  The values of $\ln {\mathrm Im}
\beta_H^{(1)}$ are shifted to
\begin{equation}
\label{eq:shift_imbetah}
\ln {\mathrm Im}\beta_H^{(1)} \rightarrow \ln {\mathrm Im}\beta_H^{(1)}
 \ - \ln (c_1 c_2)
\end{equation}
in order to use the results of both gauge couplings for the fit
(Fig.~\ref{fig:log_im_vs_log_L}).  The main shift $\ln c_1$ is derived from
Eq.~(\ref{eq:zero_as_fct_of_rho2}) assuming that the continuum condensate jump
is already independent of $a$
\begin{equation}
c_1=  \frac{\beta_{Hc}^2}{\beta_G^2 (1+2\beta_{Rc})} \ .
\label{eq:c1}
\end{equation}
In the logarithmic shift we have used the real--valued finite volume couplings
${\mathrm Re} \beta_H^{(1)}$ and $\beta_{R}({\mathrm Re}\beta_H^{(1)})$.  A
small extra shift $\ln c_2$ (with $c_2$ between 1.028 and 1.095 in the used
range of $\lambda_3/g_3^2$) has been added to correct the eventual imprecision
of the used equation.  It has been adjusted in a way to provide a minimal
$\chi^2$ for all (reweighted) data at given $\lambda_3/g_3^2$ in the fit of
Eq.~(\ref{eq:zero_fit}).  The three rightmost data points in
Fig.~\ref{fig:log_im_vs_log_L} arises from $96^3$ data at $M_H^*=74$~GeV which
are reweighted to $M_H^*=72$ and 76~GeV, too.

\begin{figure}
  \centering 
  \epsfig{ file=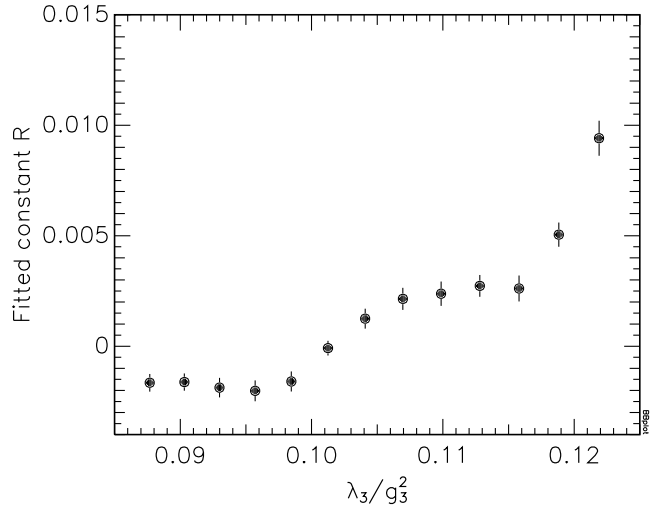, width=86mm,angle=0}
  \caption{
    Fitted distance $R$ as function of $\lambda_3/g_3^2$}
  \label{fig:zero_fit}
\end{figure}

The values of the fit constant $R$ using all lattice sizes and both gauge
couplings at fixed $\lambda_3/g_3^2$ are given in Fig.~\ref{fig:zero_fit}.
Near to the endpoint the $\chi^2$ of the fits deteriorate.  For smaller
lattice sizes the asymptotic behaviour $\propto 1/l^3$ of ${\mathrm Im}
\beta_H^{(1)}$ is still not reached. Hence the constant $R$ is found negative
as long as the phase transition is of first order. We localise the endpoint of
the transition where $R$ as a function of $\lambda_3/g_3^2$ crosses zero. This
gives the critical value
\begin{equation}
  \lambda_{3 {\mathrm crit}}/g_3^2=0.102(2)  
  \label{eq:crit_LY} 
\end{equation}
which translates into $M_{H\mathrm{crit}}^*=72.2(6)$~GeV.  

\begin{figure} 
  \centering
  \epsfig{ file=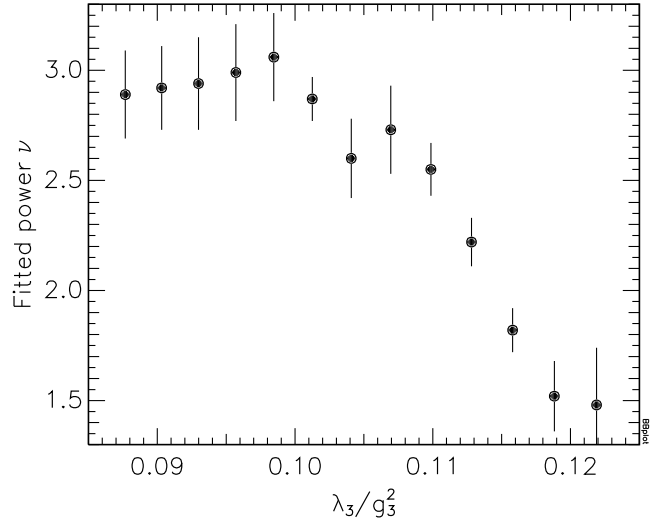, width=86mm,angle=0}
  \caption{
    Fitted power $\nu$ as function of $\lambda_3/g_3^2$}
  \label{fig:nu_fit}
\end{figure}

Restricting the fit of Eq.~(\ref{eq:zero_fit}) only to larger volumes the
expected power $\nu=3$ for the first order transition is reproduced.  This is
shown in Fig.~\ref{fig:nu_fit} where only six (seven) data points above $ \ln
(l g_3^2)=2.55$ (see Fig.~\ref{fig:log_im_vs_log_L}) are included in the fit.
For $\lambda_3/g_3^2 > 0.102$ the fit yields a power which strongly decreases.
This again indicates the change to the crossover.

This critical Higgs coupling is only slightly  below the upper bound obtained in
section~\ref{sec:latheat} from the argument of vanishing latent heat.

\section{Discussion and conclusions}
\label{sec:conclusions}
We have compared two methods which promised to give estimates for the critical
Higgs mass. We have used on one hand a criterion based on the thermodynamical
limit of Lee--Yang zeroes, requiring that the leading zero approach the real
axis in the infinite volume limit. This has lead  to the critical coupling ratio
(\ref{eq:crit_LY}).  For this purpose we had to rescale results obtained with
different values of the lattice gauge coupling, in our work $\beta_G=12$ and
$\beta_G=16$.

The criterion based on a vanishing scalar condensate tends to predict a too
high critical Higgs mass in accordance with the multihistogram interpolation.
Very near to the endpoint a two--state signal persists which is not related to
a first order phase transition. One has to use essentially larger lattices in
order to get a reliable infinite volume extrapolation.  By this technique we
have identified the upper bound (\ref{eq:l3g32_first}).

The critical temperature $T_c$ and the actual Higgs mass $m_H$ of the
underlying $4d$ theory corresponding to the endpoint of the first order
transition can be calculated using the relations in Sec.~2 of \cite{wirNP97}.
These quantities are listed in Table~\ref{tab:latheat} using the lattice
couplings $\beta_G=12$ and $\beta_{Hc}=.3437161$ at the critical continuum
coupling ratio (\ref{eq:crit_LY}) as derived from the Lee--Yang zeroes
analysis.  Additionally, the four dimensional $\overline{\mathrm MS}$ running
coupling $g^2(m_W)$ is given.  All quantities are calculated for the two cases
of the $4d$ $SU(2)$--Higgs theory, without fermions and including the top
quark.

The apparent two--state signal for $\rho^2$ near or at the endpoint is
misleading and cannot be an indicator of a first order phase transition.  The
reason is that the correlation length grows to the size of the system being
simulated. At $M_H^*=70$~GeV, for instance, these two scales can be safely
separated from each other \cite{wirNP97}. When the transition becomes
increasingly weak the situation will change rapidly.  In order to measure the
correlation length of the competing phases one would have to take some care.
One should carefully monitor the tunnelling of the system in order to measure
the correlation functions of the pure phases, respectively.  We have
successfully applied such a procedure at $M_H^*=70$~GeV. For the weaker
transitions at higher Higgs mass this becomes increasingly difficult.
Therefore we have restricted our attention exclusively to bulk variables.  At
the critical endpoint one expects the correlation length to diverge.

Our result for $\lambda_{3{\mathrm crit}}/g_3^2$ is not so far from the result
by Karsch et al. \cite{karsch96} who have obtained (in our notation)
\begin{equation}
  \label{eq:karsch}
  \lambda_{3 {\mathrm crit}}/g_3^2=0.0951(16), \ \ {\mathrm at}  \ \ \beta_G=9 \  
\end{equation}
analysing lattices with an extent $\ln (l g_3^2) \le 3.06 $.  The remaining
difference between (\ref{eq:karsch}) and (\ref{eq:crit_LY}) can comfortably be
explained by the fact that we come nearer to the continuum limit.

It might be instructive to transform our results to a $4d$ $SU(2)$--Higgs
model at a larger $\mathrm \overline {MS}$ running gauge coupling.  Usually in
$4d$ simulations, the bare coupling $g^2=0.5$ is used.  The measured
renormalised $4d$ gauge coupling does not seem to change significantly with
the Higgs mass in the so far reported region from 18 to 49~GeV
\cite{desy,bunk94} and varies from 0.56 to 0.59. We expect that this coupling
remains within this range at larger Higgs mass, too.  Since a perturbative
calculation is missing we assume here, following Refs. \cite{laine,interface},
that the measured renormalised gauge coupling roughly corresponds to the
$\overline{\mathrm{MS}}$ running coupling.

For definiteness we choose $g^2(m_W)=0.58$ and take $\lambda_3/g_3^2=0.102$.
We obtain the critical Higgs mass $m_H=65.2$~GeV and the corresponding
transition temperature $T_c=129.6$~GeV. This is noticeably smaller than the
critical Higgs mass estimated in Ref.~\cite{aoki} which is the only $4d$
result so far available.

At weakly first order transitions, the $3d$ effective theory seems to describe
the transition parameters of the $4d$ model reasonably well
\cite{laine,interface}.  Concerning the apparent first order nature of the
transition at $m_H \ge 67$~GeV in the $4d$ approach, there is reason for
doubts because of the very coarse discretisation with $N_t=2$ temporal steps.

\section*{Acknowledgements}

E.M.~I. is supported by the DFG under grant Mu932/3-4.  The use of the
DFG--Quadrics QH2 parallel computer in Bielefeld and the Q4 parallel computer
in the HLRZ J\"ulich is acknowledged.  Additionally, we thank the council of
HLRZ J\"ulich for providing CRAY-T90 resources.  M.~G. and A.~S. are grateful
to C.~Borgs on a discussion about Lee--Yang zeroes.

\newpage

\newpage
\begin{table}[]
    \begin{tabular}{ccclr|ccclr}
      $ M_H^*$    &
      $ \beta_G $ &
      $ L$        &
      $ \beta_H $ &
      sweeps      &
      $ M_H^*$    &
      $ \beta_G $ &
      $ L$        &
      $ \beta_H $ &
      sweeps\\
      \tableline
      70 & 12 &30 &0.343480  & 30000 & 74 & 12 &64 &0.343848  & 20000 \\
      70 & 12 &30 &0.343540  & 50000 & 74 & 12 &64 &0.343850  & 25000 \\    
      70 & 12 &30 &0.343600  & 40000 & 74 & 12 &64 &0.343852  & 30000 \\    
      70 & 12 &40 &0.343540  & 20000 & 74 & 12 &80 &0.3438486 & 40000 \\    
      70 & 12 &40 &0.343560  & 20000 & 74 & 12 &96 &0.3438486 & 40000 \\   
      70 & 12 &48 &0.343440  & 75000 & 76 & 12 &30 &0.343980  & 20000\\    
      70 & 12 &48 &0.343520  & 40000 & 76 & 12 &30 &0.344000  & 80000\\    
      70 & 12 &48 &0.343540  & 80000 & 76 & 12 &30 &0.344040  & 20000\\    
      70 & 12 &48 &0.343544  &120000 & 76 & 12 &40 &0.343990  & 20000\\    
      70 & 12 &48 &0.343546  & 20000 & 76 & 12 &40 &0.344000  & 30000\\     
      70 & 12 &48 &0.343548  &120000 & 76 & 12 &40 &0.344020  & 20000\\    
      70 & 12 &48 &0.343560  & 40000 & 76 & 12 &48 &0.343994  & 25000\\    
      70 & 12 &48 &0.343580  &110000 & 76 & 12 &48 &0.344000  & 35000\\    
      70 & 12 &64 &0.343546  & 90000 & 76 & 12 &48 &0.344006  & 35000\\    
      70 & 12 &64 &0.343548  &120000 & 76 & 12 &48 &0.344012  & 10000\\    
      70 & 12 &64 &0.343549  & 20000 & 76 & 12 &64 &0.344000  & 40000\\    
      70 & 12 &64 &0.343550  &100000 & 76 & 12 &64 &0.344006  & 40000\\    
      70 & 12 &80 &0.343546  & 40000 & 76 & 12 &80 &0.344002  & 20000\\    
      70 & 16 &32 &0.340780  & 40000 & 76 & 12 &80 &0.344002  & 40000\\     
      70 & 16 &32 &0.340800  & 40000 & 76 & 12 &80 &0.344006  & 25000\\    
      70 & 16 &32 &0.340820  & 40000 & 76 & 16 &32 &0.341100  & 20000\\    
      70 & 16 &40 &0.340780  & 40000 & 76 & 16 &32 &0.341120  & 40000\\    
      70 & 16 &40 &0.340800  &100000 & 76 & 16 &32 &0.341140  & 20000\\    
      70 & 16 &40 &0.340820  & 40000 & 76 & 16 &40 &0.341120  & 30000\\    
      70 & 16 &48 &0.340700  & 45000 & 76 & 16 &40 &0.341124  & 30000\\    
      70 & 16 &48 &0.340780  & 45000 & 76 & 16 &40 &0.341130  & 20000\\     
      70 & 16 &48 &0.340800  & 90000 & 76 & 16 &48 &0.341124  & 35000\\    
      70 & 16 &48 &0.340820  & 45000 & 76 & 16 &48 &0.341128  & 20000\\    
      70 & 16 &64 &0.340796  & 40000 & 76 & 16 &64 &0.341124  & 40000\\
      70 & 16 &64 &0.340800  & 80000 & 76 & 16 &64 &0.341128  & 20000\\ 
      70 & 16 &64 &0.340804  & 40000 & 76 & 16 &80 &0.341126  & 20000\\
      70 & 16 &80 &0.340802  & 30000 & 76 & 16 &80 &0.341130  & 30000\\
      74 & 12 &48 &0.343850  & 40000 & 80 & 16 &80 &0.341360  & 40000\\
    \end{tabular}
  \caption{Statistics}
  \label{tab:statistics}
\end{table}

\begin{table}[htb]
    \begin{tabular}{ccc}
      &&\\[-2ex]
      $m_H/$GeV  & $T_c$/GeV & $g^2(m_W)$ \\[+0.5ex]
      \tableline
      67.0(8)  & 154.8(2.6)& 0.423      \\
      72.4(9)  & 110.0(1.5)  & 0.429     \\
    \end{tabular}
    \caption{
      Some quantities at $\lambda_{3{\mathrm crit}}/g_3^2=0.102$ 
      ($M_{H\mathrm{crit}}^*=72.2$~GeV);
      upper row without fermions, lower  including top}
    \label{tab:latheat}
\end{table}
\end{document}